\documentclass[aps,pre,showpacs,superscriptaddress,eqsecnum]{revtex4-1}

\usepackage{graphicx}
\usepackage{dcolumn}
\usepackage{bm}
\usepackage{combelow}

\usepackage[utf8]{inputenc}
\usepackage[T1]{fontenc}
\usepackage{mathptmx} 
\usepackage{xcolor}
\usepackage{mathrsfs}
\usepackage{mathtools}

\begin{document}

\title{Implicit-Explicit Finite-Difference Lattice Boltzmann\\Model with Varying Adiabatic Index}

\author{\cb{S}tefan T. Kis} 
 \email{stefan.kis@e-uvt.ro}
\author{Victor E. Ambru\cb{s}}%
 \email[Corresponding author: ]{victor.ambrus@e-uvt.ro}
\affiliation{
  West University of Timi\cb{s}oara, 4 Vasile P\^{a}rvan Blvd., 300223, Timi\cb{s}oara, Romania.
}

\date{\today} 

\begin{abstract}
The perfect fluid limit can be obtained from the Boltzmann equation in the limit of vanishing Knudsen number. By treating the collision term in an implicit manner, the implicit-explicit (IMEX) time stepping scheme allows this limit to be achieved at finite values of the time step. We consider the $9$th order monotonicity-preserving (MP-9) scheme to implement the advection, which is treated explicitly in the IMEX approach. We reduce the computational costs using reduced distribution functions, which also permits the adiabatic index to be varied. We validate the capabilities of our model by considering the propagation of shock waves in one-dimensional and two-dimensional setups.
\end{abstract}

\maketitle

\section{\label{sec:intro}Introduction}

The development and propagation of shock waves in fluids has 
become a problem of wide interest to the computational fluid 
dynamics (CFD) community since at least a century ago. The two 
main features that make this problem so complex are its strongly 
nonlinear character and the presence of discontinuities. Despite 
these difficulties, analytic solutions can be derived in the 
limit of an inviscid (perfect) fluid, obeying the Euler 
equations, but only in some very simple cases (e.g., the Sod 
shock tube setup \cite{SOD1978}). For more complex fluids and/or 
setups, one must rely on numerical simulations for the study 
of shock wave phenomena. 

The apparent discontinuities in flows with shocks can persist only down 
to the molecular scale of the particle mean free path $\lambda$,
where the fluid is far from thermal equilibrium and the Navier-Stokes (NS) equations lose applicability. At this level, the flow can be described using the Boltzmann equation, which governs the evolution of the particle distribution function $f$. 
Since the numerical evaluation of the Boltzmann collision integral is computationally expensive \cite{Mieussens2000}, we consider the Bhatnagar-Gross-Krook (BGK) model \cite{BGK54} for the collision term. 
We discretize the velocity space discretization following the lattice Boltzmann (LB) approach, which is designed to recover the 
NS equations starting from the Boltzmann-BGK equation 
\cite{Succi2018}. In this paper, we employ the finite difference LB (FDLB) method, in which the time stepping and advection are implemented using finite differences techniques \cite{ambrus19chap}.

For the nearly-inviscid regime, the relaxation time $\tau$ of the 
BGK model must be decreased to very low values. Explicit time
solvers become unstable if the time step $\delta t$ exceeds $\tau$.
Time steps higher than $\tau$ can be employed by treating the 
collision term implicitly. Since a fully implicit treatment 
of the Boltzmann-BGK equation is expensive, we employ the Implicit-Explicit 
(IMEX) approach, which allows the advection part to remain explicit 
\cite{Pareschi2005}. An FDLB IMEX implementation was introduced 
in Ref.~\cite{Wang07} in conjunction with the fifth order 
Weighted Essentially Non-Oscillatory (WENO-5) scheme, successfully 
employing $\delta t \simeq  4.5 \times 10^5 \tau$. 
The model presented in 
Ref.~\cite{Wang07} is based on the BGK model for monatomic gases
and is therefore restricted to fluids with adiabatic index 
$\gamma = 5/3$. 

In this paper, we extend the work of Ref.~\cite{Wang07} to the case of arbitrary $\gamma$ (we focus on diatomic molecules with $\gamma = 7/5$), by using a pair of {\it reduced} distribution functions, following the procedure described in Ref.~\cite{Guo15}. We employ the 9th order monotonicity-preserving (MP-9) scheme \cite{Suresh1997} and demonstrate the capabilities of our implementation by considering the one-dimensional Sod shock tube problem \cite{SOD1978} and the 2D Riemann problem introduced in Ref.~\cite{lax98}. 

\section{Lattice Boltzmann algorithm}

The LB method starts from the Boltzmann equation with the 
BGK approximation for the collision term \cite{BGK54}: 
\begin{align}\label{eq:BoltzmannBGK}
    \partial_t f + \frac{\bm{p}}{m} \cdot \nabla f &= J[f] \simeq - \frac{1}{\tau}\left( f - f^{\,(eq)}\right), & f^{\,(eq)} &= \frac{n}{(2\pi m k_B T)^{(3+K)/2}} \exp{\left( - \frac{\bm{\xi}^2 + \bm{\eta}^2 + \bm{\zeta}^2}{2m k_B T} \right)},
\end{align}
where the relaxation time $\tau$ is taken to be constant. 
In Eq.~\eqref{eq:BoltzmannBGK}, $f^{\,(eq)}$ is the Maxwell-Boltzmann distribution, while $n$, $\bm{u}$ and $T$ are the particle number density, fluid velocity and temperature $T$.
The microscopic degrees of freedom (DOFs) are split into three categories: the first $d$ ($1 \le d \le D = 3$) DOFs represent coordinate directions along which the fluid is non-homogeneous and are denoted using $p_\alpha$, with $\xi_\alpha = p_\alpha - mu_\alpha$ being the peculiar momentum  ($1 \le \alpha \le d$, $\bm{\xi}^2 \equiv \xi_1^2 + \dots \xi_d^2$). The other $D - d$ DOFs, $\eta_i$ ($d < i \le D$, $\bm{\eta}^2 = \eta_{d+1}^2 + \dots \eta_D^2$), correspond to directions along which the fluid is homogeneous and at rest. We also consider $K$ internal DOFs, $\zeta_a$ ($1 \le a \le K$, $\bm{\zeta}^2 = \zeta_1^2 + \dots \zeta_K^2$) \cite{Guo15}. 

Since the dynamics along the $\eta_i$ and $\zeta_a$ directions is trivial, it is convenient to integrate out these DOFs and to describe the system using two reduced distributions $\phi$ and $\chi$ and their corresponding equilibria \cite{Guo15,ambrus19chap}:
\begin{equation}
 \begin{pmatrix} 
  \phi \\ \chi 
 \end{pmatrix} = \int d\bm{\eta} \, d\bm{\zeta} \, \begin{pmatrix} 1 \\ (\bm{\eta}^2 + \bm{\zeta}^2)/m \end{pmatrix} f, \qquad
 \phi^{\,(eq)} = n \prod_{\alpha = 1}^d g_\alpha, \qquad 
 g_\alpha \equiv g(p_\alpha, u_\alpha, T) = 
 \frac{\exp( - \xi_\alpha^2/2 m k_B T)}
 {\sqrt{2\pi m k_B T}},
 \label{eq:galpha}
\end{equation}
while $\chi^{\,(eq)} = (K+D-d) k_B T \phi^{\,(eq)}$. 
The equations satisfied by $\phi$ and $\chi$ can be obtained 
from Eq.~\eqref{eq:BoltzmannBGK},
while $n$, $\bm{u}$ and $T$ are obtained 
as moments of $\phi$ and $\chi$ with respect to $\bm{p}$:
\begin{equation}
 \partial_t 
    \begin{pmatrix} 
     \phi \\ \chi 
    \end{pmatrix} + \frac{\bm{p}}{m} \cdot \nabla
    \begin{pmatrix}  
     \phi \\ \chi 
    \end{pmatrix} = -\frac{1}{\tau} 
    \begin{pmatrix} 
     \phi - \phi^{\,(eq)} \\ 
     \chi - \chi^{\,(eq)}
    \end{pmatrix}, \qquad
    \begin{pmatrix} 
     n \\ \rho \bm{u} \\ \frac{D+K}{2} n k_B T
    \end{pmatrix} 
 = \int d\bm{p} \, 
 \left[ \phi 
 \begin{pmatrix} 1 \\ \bm{p} \\ \bm{\xi}^{2}/2m
 \end{pmatrix} + \chi
 \begin{pmatrix} 0 \\ 0 \\ 1/2 \end{pmatrix} \right].
\label{eq:Moments}
\end{equation}

The non-dimensionalization considered in this paper is based on reference quantities, which we introduce as follows. The reference length $L$ is taken to be the linear size of the domain. Denoting the reference density and temperature using $n_{\rm ref}$ and $T_{\rm ref}$, the reference speed, momentum and pressure are $c_{\rm ref} = \sqrt{k_B T_{\rm ref} / m}$, $p_{\rm ref} = m c_{\rm ref}$ and $P_{\rm ref} = n_{\rm ref} k_B T_{\rm ref}$. The reference time is $t_{\rm ref} = L_{\rm ref} / c_{\rm ref}$. 

Discretizing the velocity space using the Gauss-Hermite quadrature, the integrals in Eq.~\eqref{eq:Moments} are replaced with quadrature sums:
\begin{equation}
 \begin{pmatrix} 
  n \\ \rho \bm{u} \\ \frac{D+K}{2} n T
 \end{pmatrix} \equiv \sum_{\bm{k}}
 \phi_{\bm{k}} 
 \begin{pmatrix} 1 \\ \bm{p}_{\bm{k}} \\ \bm{\xi}^{2}_{\bm{k}}/2m
 \end{pmatrix} + 
 \sum_{\bm{q}}
 \chi_{\bm{q}}
 \begin{pmatrix} 0 \\ 0 \\ 1/2 \end{pmatrix}
  = \sum_{\bm{k}}
 \phi^{\,(eq)}_{\bm{k}} 
 \begin{pmatrix} 1 \\ \bm{p}_{\bm{k}} \\ \bm{\xi}^{2}_{\bm{k}}/2m
 \end{pmatrix} + 
 \sum_{\bm{q}}
 \chi^{\,(eq)}_{\bm{q}}
 \begin{pmatrix} 0 \\ 0 \\ 1/2 \end{pmatrix},
 \label{eq:Moments_discrete}
\end{equation}
On each axis $\alpha$ ($1 \le \alpha \le d$) of the momentum space, $Q_\phi$ ($1 \le k_\alpha \le Q_\phi$) and $Q_\chi$ ($1 \le q_\alpha \le Q_\chi$) quadrature points are employed, which are obtained as the roots of the Hermite polynomials $H_{Q_{\phi}}(p_{\alpha,k_\alpha})$ and 
$H_{Q_{\chi}}(p_{\alpha,q_\alpha})$. The total number of momentum vectors employed is $Q_\phi^d$ and $Q_\chi^d$ for $\phi$ and $\chi$, respectively, which are collectively denoted using $\bm{k} = (k_1, \dots k_d)$ and $\bm{q} = (q_1, \dots q_d)$. The corresponding discrete populations $\phi_{\bm{k}}$ and $\chi_{\bm{q}}$ are related to $\phi$ and $\chi$ through:
\begin{equation}
 \phi_{\bm{k}} = \frac{w_{k_1}^H(Q_\phi) \times \cdots
 w_{k_d}^H(Q_\phi)}{\exp(-\bm{p}_{\bm{k}}^2/2)}
 \phi(\bm{p}_{\bm{k}}),
 \qquad 
 \chi_{\bm{q}} = \frac{w_{q_1}^H(Q_\chi) \times \cdots
 w_{q_d}^H(Q_\chi)}{\exp(-\bm{p}_{\bm{q}}^2 / 2)}
 \chi(\bm{p}_{\bm{q}}),\qquad 
 w_k^H(Q) =\frac{Q!}{[H_{Q+1}(p_k)]^2},
\end{equation}
where the quadrature weights $w_k^H(Q)$ were incorporated for convenience in the definition of the discrete populations.

The discrete equilibria, $\phi_{\bm{k}}^{\,(eq)}$ and
$\chi_{\bm{q}}^{\,(eq)}$, are constructed such that the equality in Eq.~\eqref{eq:Moments_discrete} is exact. 
This is achieved by truncating 
the expansion of $g_\alpha$ \eqref{eq:galpha} with respect to the Hermite polynomials at a finite order 
$Q_{\phi/\chi} - 1$
\cite{ambrus19chap,Ambrus2016}:
\begin{align}
 \phi^{\,(eq)}_{\bm{k}} = n \prod^{d}_{\alpha = 1} g_{\alpha, k_{\alpha}}^{Q_\phi-1}, \qquad 
 g_{\alpha, k_{\alpha}}^{Q_\phi-1} = w^{H}_{k_{\alpha}}(Q_{\phi}) \sum^{Q_\phi-1}_{\ell = 0} \frac{1}{\ell!}  H_{\ell} (p_{k_{\alpha}}) \sum_{s = 0}^{\lfloor \ell/2 \rfloor} \frac{\ell !}{2^{s} \, s! \left(\ell - 2s \right)!} (mT- 1)^{s} (m u_{\alpha})^{\ell - 2s},
 \label{eq:gExpanded}
\end{align}
and similarly for $\chi_{\bm{q}}^{\,(eq)}$ with 
$k_\alpha$ and $Q_\phi$ replaced by $q_\alpha$ and $Q_\chi$.
The above expansion ensure the exact recovery of the moments up to orders $Q_\phi - 1$ and $Q_\chi -1$ with respect to each axis of $\phi$ and $\chi$ via the quadrature sums in Eq.~\eqref{eq:Moments_discrete}. 

\section{Implicit-explicit Runge-Kutta time-stepping scheme}

\begin{table}
\begin{center}
\begin{tabular}{cccc}
Explicit: \qquad & \qquad Implicit: \qquad \qquad & \qquad \qquad Explicit: \qquad & \qquad Implicit: \\
\begin{tabular}{c|c}
 $\widetilde{c}$ & $\widetilde{A}$ \\\hline
 & $\widetilde{w}^T$ 
\end{tabular}, \qquad &
\qquad \begin{tabular}{c|c}
 $c$ & $A$ \\\hline
 & $w^T$ 
\end{tabular}, \qquad \qquad & \qquad \qquad 
{\small 
\begin{tabular}{c|cccc}
 $0$ & $0$ & $0$ & $0$ & $0$\\
 $0$ & $0$ & $0$ & $0$ & $0$\\
 $1$ & $0$ & $1$ & $0$ & $0$\\
 $1/2$ & $0$ & $1/4$ & $1/4$ & $0$ \\\hline
 & $0$ & $1/6$ & $1/6$ & $2/3$
\end{tabular}, } \qquad & \qquad 
{\small 
\begin{tabular}{c|cccc}
 $\alpha$ & $\alpha$ & $0$ & $0$ & $0$ \\
 $0$ & $-\alpha$ & $\alpha$ & $0$ & $0$ \\
 $1$ & $0$ & $1-\alpha$ & $\alpha$ & $0$ \\
 $1/2$ & $\beta$ & $\eta$ & $1/2-\beta-\eta-\alpha$ & $\alpha$ \\\hline
 & $0$ & $1/6$ & $1/6$ & $2/3$
\end{tabular},}
\end{tabular}
\end{center}
\caption{ Butcher tableaux structure for the IMEX-SSP3(4,3,3) 
scheme, containing separately the coefficients for the Explicit (left) and Implicit (right) parts of the algorithm. The notations 
$\widetilde{c}_i = \sum^{i-1}_{j=1} \widetilde{a}_{ij}$ and $c_{i} = \sum^{i}_{j=1} a_{ij}$ give the time moments at the intermediate stages $i$, while $\alpha = 0.24169426078821$, $\beta = 0.06042356519705$ and $\eta = 0.12915286960590$ 
are constants \cite{Pareschi2005,Wang07}.}
\label{tab:IMEX}
\end{table}

Following Ref.~\cite{Wang07}, we employ the $3^{\rm{rd}}$ order strong stability preserving implicit-explicit (IMEX) Runge-Kutta scheme, denoted IMEX-SSP3(4,3,3) in Ref.~\cite{Pareschi2005}, for the time evolution. This scheme treats implicitly the collision part of the Boltzmann equation, remaining stable when $\delta t > \tau$. 
Considering that $F_n \equiv F(t_n)$, where $F \in \{\phi,\chi\}$, 
is known at time $t_n$, its value $F_{n+1}$ at $t_{n+1} = t + \delta t$ can be obtained using $r$ intermediate stages, $1 \leq i,j \leq r$, summarized as follows \cite{Wang07}:
\begin{equation}
    F^{\,(i)} = F^{\,n} - \delta t \sum^{i-1}_{j=1} \widetilde{a}_{ij} \frac{\bm{p}}{m} \cdot \nabla F^{\,(j)} - \sum_{j=1}^{i} a_{ij} \delta F^{\,(j)},\qquad
    F^{\,n+1} = F^{\,n} - \delta t \sum^{r}_{i=1} \widetilde{w}_{i} \frac{\bm{p}}{m} \cdot \nabla F^{\,(i)} - \sum_{i=1}^{r} w_i \delta F^{\,(i)},
    \label{eq:IMEX}
\end{equation}
where $\delta F^{\,(i)} \equiv \frac{\delta t}{\tau} [F^{\,(i)} - F^{\,(i)}_{\,(eq)}]$. The coefficients $\widetilde{a}_{ij}$, $a_{ij}$, $\widetilde{w}_{i}$ and $w_{i}$ are summarized via Butcher tableaux in Table~\ref{tab:IMEX}. 

The only unknown quantity when computing $F^{\,(i)}$ is $F^{\,(i)}_{\,(eq)}$. 
Generally, the construction of $F^{\,(i)}_{\,(eq)}$ requires knowledge of $n^{\,(i)}$, $\bm{u}^{\,(i)}$ and $T^{\,(i)}$.
These quantities can be obtained by multiplying Eq.~\eqref{eq:IMEX} by the collision invariants $\psi \in \{1, \bm{p}, (\bm{p}^2+\bm{\eta}^2 + \bm{\zeta}^2)/2m\}$ and integrating over the microscopic DOFs. In general, $\delta F^{\,(i)}$ make vanishing contributions to these integrals. For the first stage, we have
$F^{\,(1)} = F^n + \alpha \delta F^{\,(1)}$. It can be seen that $\rho^{\,(1)} = \rho^n$, $\bm{u}^{\,(1)} = \bm{u}^{\,n}$ and $T^{\,(1)} = T^{\,n}$, such that $F^{\,(1)}_{\,(eq)} = F^n_{\,(eq)}$. A similar argument holds for the second stage, such that:
\begin{equation}
 F^{\,(1)} = \frac{F^{\,n} + \alpha \delta t F^{\,n}_{\,(eq)} / \tau}
 {1 + \alpha \delta t / \tau}, \qquad 
 F^{\,(2)} = \frac{F^{\,n} + \alpha \delta t F^{\,(1)} / \tau}
 {1 + \alpha \delta t / \tau}.
\end{equation}
The third and fourth stages can be computed starting from:
\begin{align}
 \left(1 + \frac{\alpha \delta t}{\tau}\right) F^{\,(3)} =&
 F^{\,n} - \Delta F^{\,(2)} +  
 \delta F^{\,(2)} + \frac{\alpha \delta t}{\tau} F^{\,(3)}_{\,(eq)}, \label{eq:IMEX_s34}\\
 \left(1 + \frac{\alpha \delta t}{\tau}\right) F^{\,(4)} =&
 F^{\,n} - \frac{\Delta F^{\,(2)} + \Delta F^{\,(3)}}{4} -
 \beta \delta F^{\,(1)} -
 \left(\eta - \frac{1}{4}\right) \delta F^{\,(2)} -
 \left(\frac{1}{4} - \alpha - \beta -\eta\right) 
 \delta F^{\,(3)} + \frac{\alpha\delta t}{\tau} F^{\,(4)}_{\,(eq)},\nonumber
\end{align}
where $\Delta F^{\,(i)} = F^{\,(i)} - \widetilde{F}^{\,(i)}$,
while $\widetilde{F}^{\,(i)} = F^{\,(i)} - \delta t \frac{\bm{p}}{m} \cdot \nabla F^{\,(i)} - \delta F^{\,(i)}$ is the result of applying an Euler evolution step to $F^{\,(i)}$. 
$F^{(3)}_{\,(eq)}$ and $F^{\,(4)}_{\,(eq)}$ can be constructed by noting that $\rho^{\,(3)} = \widetilde{\rho}^{\,(2)}$, $\bm{u}^{\,(3)} = \widetilde{\bm{u}}^{\,(2)}$ and $T^{\,(3)} = \widetilde{T}^{\,(2)}$, while
\begin{gather}
 \rho^{\,(4)} = \frac{3}{4} \rho^{\,n} + 
 \frac{1}{4} \widetilde{\rho}^{\,(3)}, \qquad 
 \rho^{\,(4)} \widetilde{\bm{u}}^{\,(4)} =
 \frac{3}{4} \rho^n \bm{u}^n + \frac{1}{4} \widetilde{\rho}^{\,(3)} \widetilde{\bm{u}}^{\,(3)}, \nonumber\\
 \frac{D + K}{2} \rho^{\,(4)} T^{\,(4)} = 
 \frac{3 \rho^{\,n} \widetilde{\rho}^{\,(3)}}
 {32 \rho^{\,(4)}} 
 [\bm{u}^{\,n} - \widetilde{\bm{u}}^{\,(3)}]^2 +
 \frac{D + K}{2} \left[\frac{3}{4} \rho^{\,n} T^{\,n}
 + \frac{1}{4} \widetilde{\rho}^{\,(3)} \widetilde{T}^{\,(3)}\right].
\end{gather}
Afterwards, $F^{\,(3)}$ and $F^{\,(4)}$ are computed from Eq.~\eqref{eq:IMEX_s34}, while $F^{\,n+1} = F^{\,n} - \frac{1}{6} \Delta F^{\,(2)} -
 \frac{1}{6} \Delta F^{\,(3)} - \frac{2}{3} \Delta F^{\,(4)}$.

\section{Monotonicity-preserving advection scheme}

We compute the advection term appearing in the Boltzmann equation, Eq.~\eqref{eq:BoltzmannBGK}, using the 9th order monotonicity-preserving (MP-9) numerical scheme introduced in 
Ref.~\cite{Suresh1997}. For simplicity, we adopt a dimensionally-unsplit approach and consider the multi-dimensional advection separately with respect to each axis.
We further illustrate the MP-9 algorithm for the case of a one-dimensional advection problem. Considering an equidistant discretization of a domain of length $L$ using $S$ intervals centered on $x_s = x_{\rm left} + (s - \frac{1}{2}) \delta s$, where 
$\delta s = L/S$ is the grid spacing, the advection operator is approximated using a flux-based approach:
\begin{equation}
    \left(\frac{p}{m} \frac{\partial F}{\partial x}  \right)_{s} = \frac{\mathcal{F}_{s+1/2} - \mathcal{F}_{s-1/2}}{\delta s},
\end{equation}
where $F \in \{\phi, \chi\}$ is any of the reduced distributions (the velocity indices $\bm{k}$ or $\bm{q}$ are omitted for brevity). 
Since the advection part is treated explicitly in the IMEX approach,  the scheme is stable when the Courant-Friedrichs-Lewy condition, $CFL = |p| \delta t / m\delta s \leq 1$, is satisfied for all advection velocities $p/m$. This criterion imposes an upper limit for $\delta t$.

Next, we discuss the construction of the flux $\mathcal{F}_{s+1/2}$ in the MP-9 scheme. For simplicity, we only discuss the case of positive advection velocity, $p / m > 0$, in which case we have \cite{SergiuPhD,Suresh1997}:
\begin{equation}
    \mathcal{F}_{s+\frac{1}{2}} = 
    \begin{cases}
        \mathcal{F}^{L}_{s+\frac{1}{2}}, & \text{for } \left(\mathcal{F}^{L}_{s+\frac{1}{2}} - F_{s} \right)\left(\mathcal{F}^{L}_{s+\frac{1}{2}} - 
        \mathcal{F}^{MP}_{s+\frac{1}{2}}\right) < 0,\\
        \mathcal{F}^{L}_{s+\frac{1}{2}}+{\rm minmod} \left( \mathcal{F}^{min}_{s+\frac{1}{2}} - \mathcal{F}^{L}_{s+\frac{1}{2}},\mathcal{F}^{max}_{s+\frac{1}{2}} - \mathcal{F}^{L}_{s+\frac{1}{2}} \right), & \text{for } \left( \mathcal{F}^{L}_{s+\frac{1}{2}} - F_{s} \right)\left( \mathcal{F}^{L}_{s+\frac{1}{2}} - \mathcal{F}^{MP}_{s+\frac{1}{2}} \right) > 0,
    \end{cases}
\end{equation}
where ${\rm minmod}(x,y) = \frac{1}{2}[{\rm sgn}(x) + {\rm sgn}(y)] {\rm min}(|x|,|y|)$ is the flux limiter \cite{Suresh1997} and the interface values $\mathcal{F}^{L}_{s+1/2}$ are computed using: 
\begin{equation}
    \mathcal{F}^{L}_{s+1/2} = \frac{1}{2520} \left( 4 F_{s-4} - 41 F_{s-3} + 199 F_{s-2} - 641 F_{s-1} + 1879 F_{s} + 1375 F_{s+1} - 305 F_{s+2} + 55 F_{s+3} - 5 F_{s+4} \right).
\end{equation}
The MP fluxes $\mathcal{F}_{s+1/2}^{MP}$ are given by:
\begin{equation}
    \mathcal{F}^{MP}_{s+1/2} = F_{s} + {\rm{minmod}} \left[F_{s+1} - F_{s}, \alpha_{\rm MP}\left(F_{s}-F_{s-1}\right)\right],
\end{equation}
where $\alpha_{\rm MP}$ is a coefficient that has to satisfy $\alpha_{\rm MP} \geq 2$ \cite{Suresh1997}. In this work, we employ $\alpha_{\rm MP} = 4$ \cite{Suresh1997}. The minimum and maximum fluxes, $\mathcal{F}^{min}_{s+1/2}$ and $\mathcal{F}^{max}_{s+1/2}$, are defined through:
\begin{align}
    \begin{pmatrix}\mathcal{F}^{min}_{s+1/2} \\ \mathcal{F}^{max}_{s+1/2} \end{pmatrix} &= \begin{pmatrix} \rm{max} \\ \rm{min} \end{pmatrix} \left[\begin{pmatrix}\rm{min} \\ \rm{max} \end{pmatrix} \left(F_{s},F_{s+1},\mathcal{F}^{MD}_{s+1/2}\right), \begin{pmatrix} \rm{min} \\ \rm{max} \end{pmatrix} \left(F_{s},\mathcal{F}^{UL}_{s+1/2},\mathcal{F}^{LC}_{s+1/2}\right)\right].
\end{align}
The notations $\mathcal{F}^{UL}_{s+1/2}$, $\mathcal{F}^{LC}_{s+1/2}$ and $\mathcal{F}^{MD}_{s+1/2}$ stand for:
\begin{align}
    \begin{pmatrix} \mathcal{F}^{UL}_{s+1/2} \\ \mathcal{F}^{LC}_{s+1/2} \end{pmatrix} &= F_{s} + \begin{pmatrix} \alpha_{\rm MP} \\ 1/2 \end{pmatrix} \left(F_{s} - F_{s-1}\right) + \begin{bmatrix} 0 \\ \frac{4}{3} {\rm{minmod}} \left(4 C_{s} - C_{s-1}, 4 C_{s-1} - C_{s}, C_{s},C_{s-1}\right)\end{bmatrix}, \\
    \mathcal{F}^{MD}_{s+1/2} &= \frac{1}{2} \left[ F_{s} + F_{s+1} - {\rm{minmod}} \left(4 C_{s} - C_{s+1}, 4C_{s+1} - C_{s}, C_{s}, C_{s+1}\right) \right],
\end{align}
where $C_{s} = F_{s+1} - 2F_s + F_{s-1}$.

Before ending this section, we address the implementation of the above algorithm in the boundary nodes. It can be seen that in the upstream direction, information from five neighbouring nodes is required. In order to apply the MP-9 algorithm, the fluid domain is extended on either side of each coordinate axis by 5 ghost nodes. Outlet boundary conditions are imposed by populating the ghost nodes using the values of the distributions in the fluid node adjacent to the boundary, on the direction perpendicular to the boundary.

\section{Numerical results}

\begin{figure}
\centering
\begin{tabular}{cc}
\includegraphics[width=0.45\linewidth]{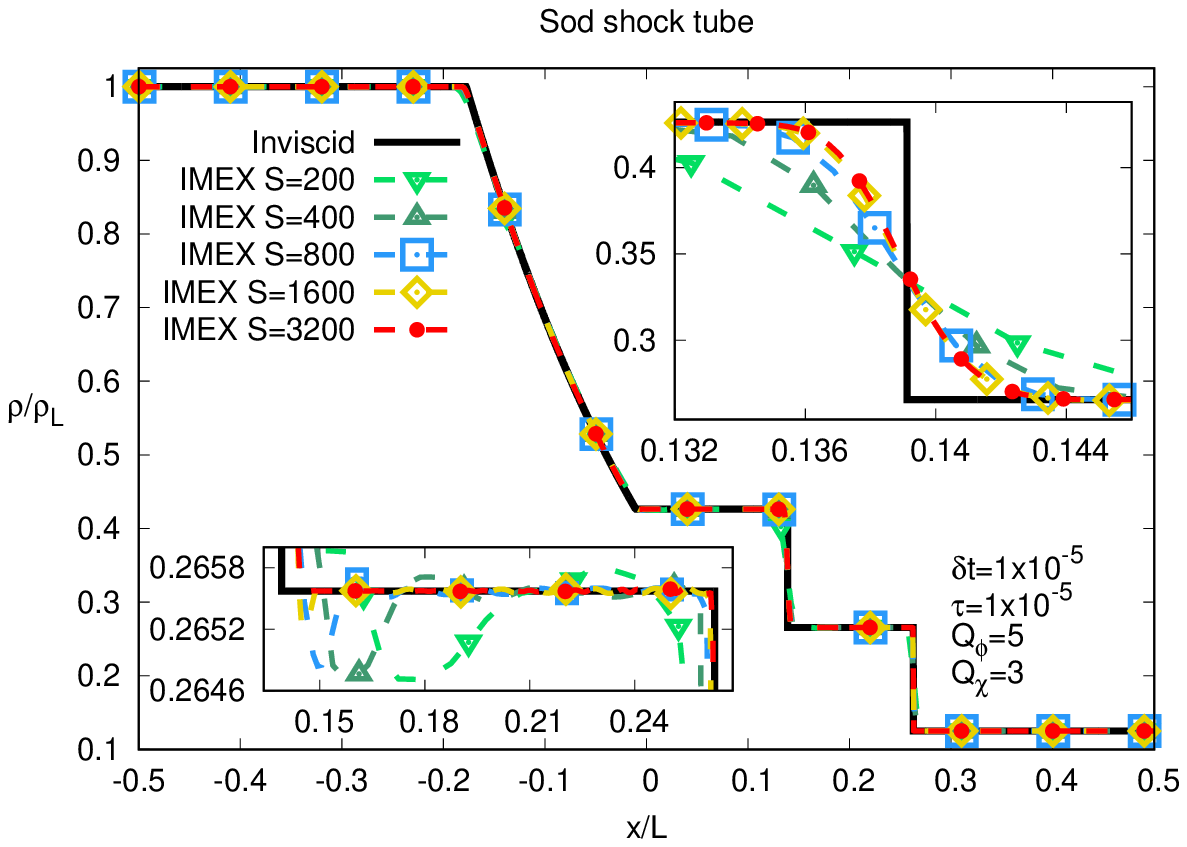} & 
\includegraphics[width=0.45\linewidth]{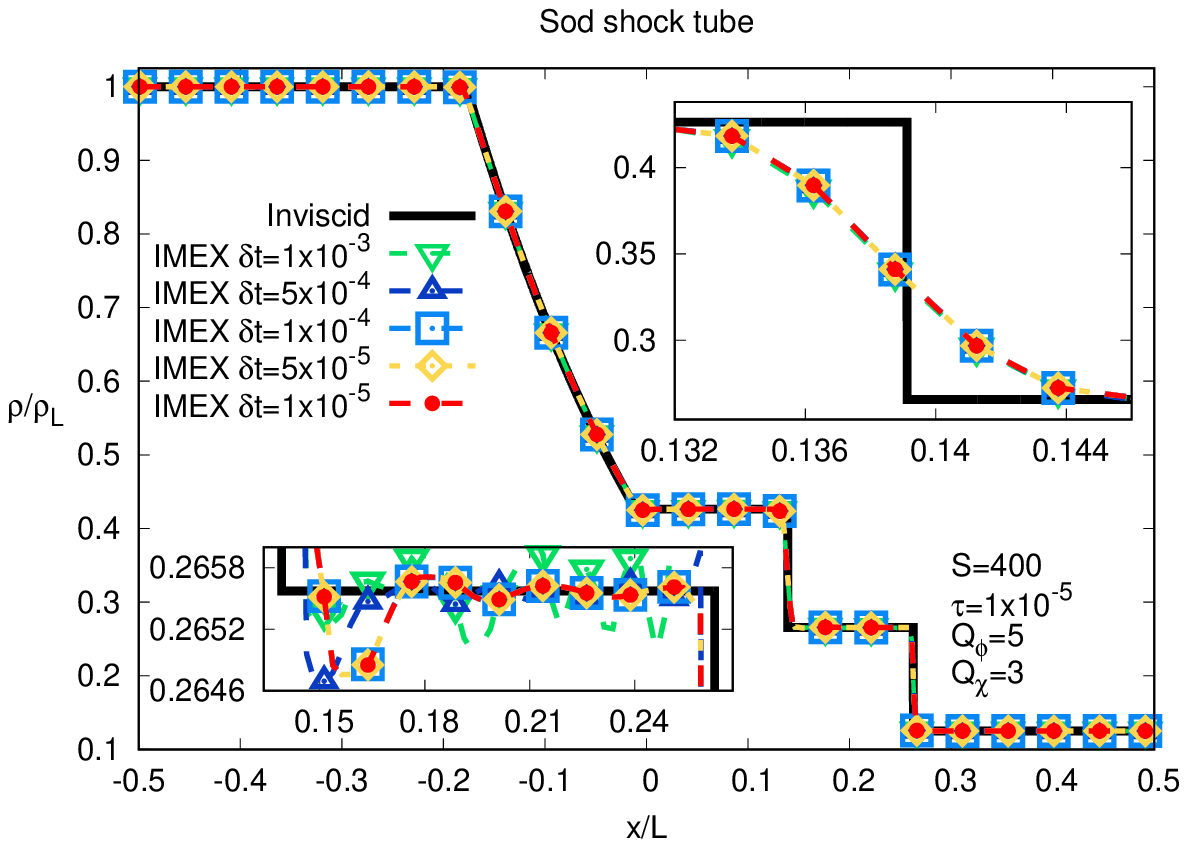}\\
(a) & (b) 
\end{tabular}

\caption{{\small Density profile at $t = 0.15$ obtained in the Sod shock 
tube setup. 
The analytic solution is shown with 
the solid black lines. 
Comparison of the numerical results obtained using the IMEX-SSP3(4,3,3) scheme for (a)
time step $\delta t = 1\times10^{-5}$ with $S = 200$, $400$, $800$, $1600$ and $3200$ nodes and
(b) $S = 400$ nodes with $\delta t = 10^{-3}$, $5 \times 10^{-4}$, $10^{-4}$, $5 \times 10^{-5}$ and $1 \times 10^{-5}$ against the analytic solution. The insets show a zoom on the CD (top right) and the plateau between the CD and SF (bottom left).}}

\label{fig:1D}
\end{figure}

\quad We first validate our implementation by considering the one-dimensional Sod shock tube problem \cite{SOD1978}. At initial time, a membrane located at $x = 0$ separates two semi-infinite domains. The fluid velocity vanishes everywhere and the fluid properties are homogeneous in each domain. The properties of the fluid to the left of the membrane are $\rho_{L} = P_{L} = T_{L} = 1$, while on the right of the membrane, $\rho_{R} = 0.125$, $P_{R}=0.1$ and $T_{R}=0.8$. At $t = 0$, the membrane is removed and a shock wave propagates from left to right. In the inviscid limit, an analytic solution can be derived \cite{SOD1978}.
This analytic solution is represented with black lines in Fig.~\ref{fig:1D}, where one can distinguish the left unperturbed region, the rarefaction wave RW, the first and second density plateaus separated by the contact discontinuity (CD), and the right unperturbed region preceeded by the shock front (SF).
Taking advantage of the homogeneity of the flow with respect to $y$ and $z$, we set $d = 1$. In addition, we set $K=2$, corresponding to two internal DOFs ($\gamma = 1.4$). The relaxation time is set to $\tau = 10^{-5}$. The quadrature orders employed in this section are $Q_\phi = 5$ and $Q_\chi = 3$, resulting in $8$ discrete populations.

In Fig.~\ref{fig:1D} (a), we investigate the effect of varying the number of nodes between $S = 200$ and $3200$, at constant time step $\delta t = 10^{-5}$. Good results are obtained. The inset in the top right shows that the CD is wider at smaller $S$, due to the numerical dissipation of the advection scheme. As $S$ is increased, a convergence trend can be observed, revealing the physical dissipation at $\tau = 10^{-5}$. On the bottom left inset, spurious fluctuations can be seen on the second plateau between the CD and SF, which are no longer visible on the scale of the inset when $S \gtrsim 1600$. Since the result corresponding to $S= 200$ presents a slight but visible deviation from the inviscid limit near the CD, we employ $S = 400$ for the rest of the paper. Fig.~\ref{fig:1D} (b) focuses on the effect of varying $\delta t$ between $10^{-3}$ and $10^{-5}$, while keeping $S= 400$ fixed.
In general, good agreement with the analytic solution is seen. The top right inset shows interestingly that the width of the CD does not appear to depend on $\delta t$. The bottom left inset shows that the $\delta t = 10^{-3}$ result presents strong oscillations on the second plateau. On the scale of this inset, it can be seen that the IMEX results with $\delta t = 5 \times 10^{-4}$ are in good agreement with those for smaller $\delta t$ and with the analytic curve, therefore we employ this value of $\delta t$ for the rest of this paper. 


\begin{figure}
\centering
\begin{tabular}{cc}
\includegraphics[width=0.45\linewidth]{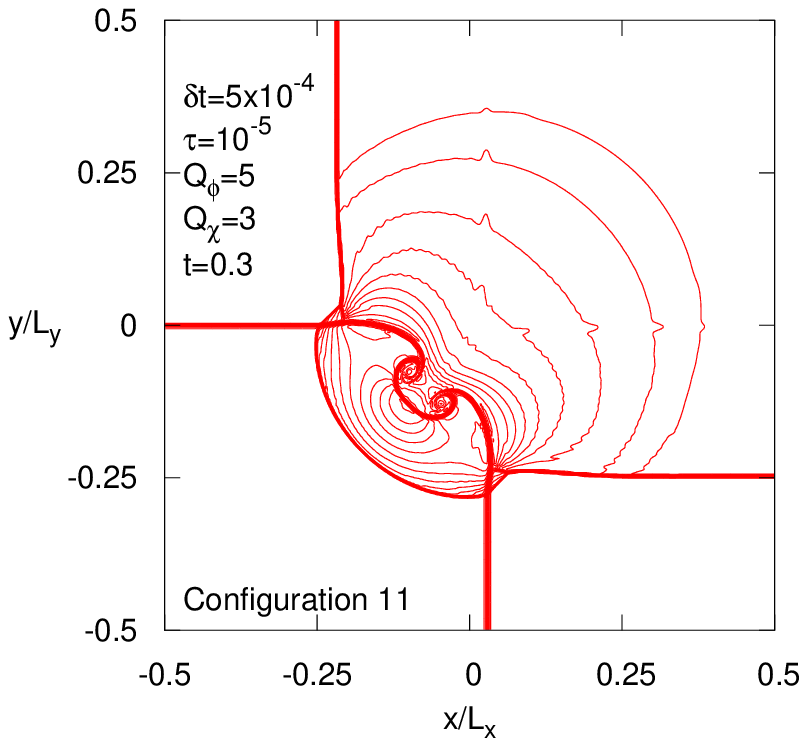} & 
\includegraphics[width=0.45\linewidth]{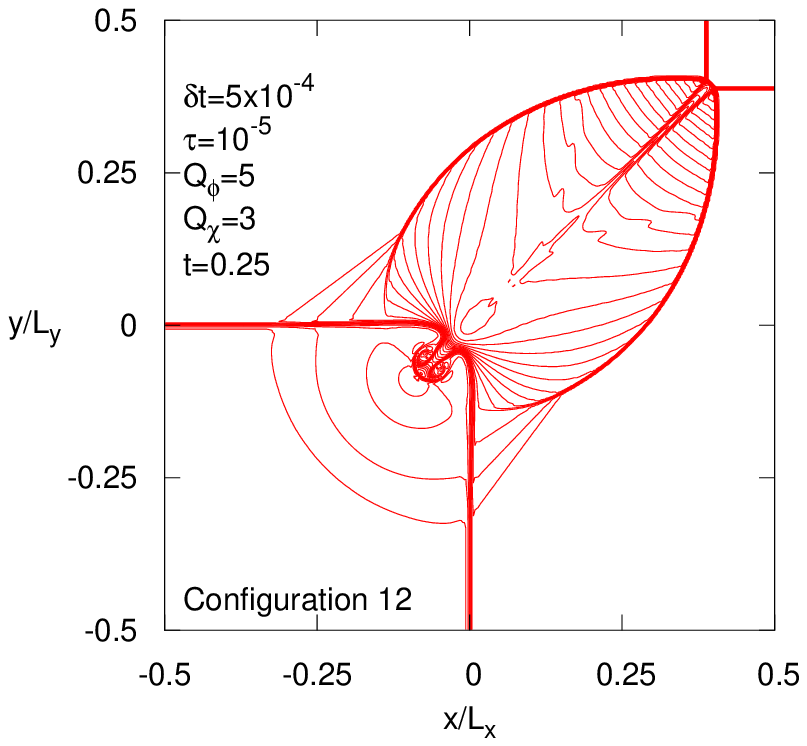}\\
(a) & (b) 
\end{tabular}
\caption{{\small Density contour plots for configurations (a) 11 and (b) 12 of the 2D Riemann problem at times $t=0.3$ and $0.25$.}}
\label{fig:2D}
\end{figure}

We now consider the two-dimensional (2D) Riemann problem, as formulated in Ref.~\cite{lax98}. The (infinite) flow domain consists of four quadrants, separated by thin membranes located along the coordinate axes. The whole setup is homogeneous with respect to the $z$ axis. Labeling the quadrants with $Q1$ (top right), $Q2$ (top left), $Q3$ (bottom left) and $Q4$ (bottom right), we consider the initial conditions corresponding to configurations 11 (C11) and 12 (C12) in Ref.~\cite{lax98}. For C11, we have $\rho_1=1$, $\rho_2 = \rho_4 = 0.5313$, $\rho_3=0.8$, $P_1=1$, $P_2=P_3=P_4=0.4$, $u_x^1 = u_x^3 = u_x^4 = 0.1$, $u_x^2=0.8276$, $u_y^1=u_y^2=u_y^3 = 0$, and $u_y^4 = 0.7276$. For C12, we have $\rho_1=0.5313$, $\rho_2 = \rho_4 = 1$, $\rho_3 = 0.8$, $P_1=0.4$, $P_2=P_3=P_4 = 1$, $u_x^1 = u_x^3 = u_x^4 = 0$, $u_x^2=0.7276$, $u_y^1=u_y^2=u_y^3=0$, and $u_y^4=0.7276$. The quadrature orders employed are $Q_\phi = 5$ and $Q_\chi = 3$, resulting in $34$ discrete populations.
It can be seen that, for both configurations, there are no discontinuities in the initial pressure and normal velocity along the interface between $Q2$ and $Q3$, as well as between $Q3$ and $Q4$. Therefore, 
slip lines (SLs) will propagate across these interfaces. SFs develop across the interface between $Q1$ and $Q2$, as well as between $Q1$ and $Q4$. In C11, the SFs propagate outwards from $Q1$, since $P_1$ and $\rho_1$ exceed those in the neighbouring quadrants. Conversely, in C12, the SFs propagate into $Q1$. 
The contour plots of the density field were obtained using an equidistant sampling of step $0.025$ from $\rho_{\rm min} = 0.52$ to $\rho_{\rm max} = 1.231$ (for C11) and $\rho_{\rm min}=0.53$ to $\rho_{\rm max} = 1.74$ (for C12). With these parameters, our scheme reproduces with good accuracy the results reported in Ref.~\cite{lax98}. 


\section{\label{sec:conc}Conclusions}

In this paper, we presented a finite difference lattice Boltzmann model for shock wave problems using the finite difference lattice Boltzmann (FDLB) method. Our implementation joins the 3rd order strong stability preserving implicit-explicit IMEX-SSP3(4,3,3) Runge-Kutta time stepping scheme, which allows the time step $\delta t$ to exceed the relaxation time $\tau$, the 9th order monotonicity-preserving MP-9 scheme for the advection, and the double distribution function approach to account for internal degrees of freedom (DOFs). Our implementation is validated in the one-dimensional Sod shock tube setup against the analytic solution and in the two-dimensional Riemann problem setup against previously published results. We obtain a $\delta t / \tau = 50$-fold acceleration compared to fully explicit schemes and we obtain a good match to the inviscid results using $\tau = 10^{-5}$ and only $400$ nodes for each axis of the simulation domain. For future work, we aim to extend the present methodology by considering higher quadrature orders for the study of the shock wave structure at the mean free path scale.

\begin{acknowledgments}
This work was supported by a grant of the Romanian Ministry of Research and Innovation, CCCDI-UEFISCDI, project number PN-III-P1-1.2-PCCDI-2017-0371, within PNCDI III.
\end{acknowledgments}
\nocite{*}
\bibliography{aipsamp}

\end{document}